\documentclass[prl,twocolumn,superscriptaddress,nofootinbib]{revtex4-1}
\usepackage[latin9]{inputenc}
\usepackage{color}
\usepackage{amsmath}
\usepackage{amssymb}
\usepackage{graphicx}
\usepackage[unicode=true,pdfusetitle,
 bookmarks=true,bookmarksnumbered=false,bookmarksopen=false,
 breaklinks=false,pdfborder={0 0 1},backref=false,colorlinks=true]
 {hyperref}
\hypersetup{
 citecolor=blue}

\makeatletter

\providecommand{\tabularnewline}{\\}

\@ifundefined{textcolor}{}
{%
  \definecolor{BLACK}{gray}{0}
  \definecolor{WHITE}{gray}{1}
  \definecolor{RED}{rgb}{1,0,0}
  \definecolor{GREEN}{rgb}{0,1,0}
  \definecolor{BLUE}{rgb}{0,0,1}
  \definecolor{CYAN}{cmyk}{1,0,0,0}
  \definecolor{MAGENTA}{cmyk}{0,1,0,0}
  \definecolor{YELLOW}{cmyk}{0,0,1,0}
}

\usepackage{braket}

\makeatother

\begin{document}
\title{\textcolor{black}{\normalsize{}Model-independent determination of
the Migdal effect via photoabsorption}}
\author{C.-P.~Liu}
\email{cpliu@mail.ndhu.edu.tw}

\affiliation{Department of Physics, National Dong Hwa University, Shoufeng, Hualien
974301, Taiwan}
\author{Chih-Pan~Wu}
\email{chih-pan.wu@umontreal.ca}

\affiliation{Département de physique, Université de Montréal, Montréal, H3C 3J7,
Canada}
\author{Hsin-Chang~Chi}
\email{hsinchang@gms.ndhu.edu.tw}

\affiliation{Department of Physics, National Dong Hwa University, Shoufeng, Hualien
974301, Taiwan}
\author{Jiunn-Wei~Chen}
\email{jwc@phys.ntu.edu.tw}

\affiliation{Department of Physics, Center for Theoretical Physics, and Leung Center
for Cosmology and Particle Astrophysics, National Taiwan University,
Taipei 10617, Taiwan}
\date{\today}
\begin{abstract}
The Migdal effect in a dark-matter-nucleus scattering extends the
direct search experiments to the sub-GeV mass region through electron
ionization with sub-keV detection thresholds. In this paper, we derive
a rigorous and model-independent \textquotedblleft Migdal-photoabsorption\textquotedblright{}
relation that links the sub-keV Migdal process to photoabsorption.
This relation is free of theoretical uncertainties as it only requires
the photoabsorption cross section as the experimental input. Validity
of this relation is explicitly checked in the case of xenon with a
state-of-the-arts atomic calculation that is well-benchmarked by experiments.
The predictions based on this relation for xenon, argon, semiconductor
silicon and germanium detectors are presented and discussed.
\end{abstract}
\maketitle

\paragraph*{Introduction.}

Direct searches for the weakly-interacting massive particle (WIMP),
one of the favorite dark matter (DM) candidates, has been making tremendous
progress in recent years: In the mass range of $m_{\chi}\sim10-100\,\textrm{GeV}$,~\footnote{We use the natural units $\hbar=c=1$ exclusively in this paper.}
the limits on its spin-independent scattering cross section off nucleon
reach down to the range of $\sigma_{n}\lesssim10^{-46}-10^{-47}\,\textrm{cm}^{2}$;
in the mass range of $1-10\,\textrm{GeV}$, the limits are also improving,
however, not as stringent as in the heavier case (see, e.g., Ref.~\citep{Zyla:2020pdg}
for a recent review). The obvious reason is a lighter WIMP has less
kinetic energy so it is less probable to generate observable nuclear
recoil (NR) events from WIMP-nucleus scattering. Ultimately, a detector's
NR threshold cuts off any sensitivity to $m_{\chi}$ below a certain
value, as seen in every WIMP exclusion plot. 

To expand a direct detector's coverage of low-mass WIMPs, or more
generically light dark matter (LDM), a recent proposal by Ibe et al.~\citep{Ibe:2017yqa}
that uses the so-called Migdal effect has attracted great interests.
This effect, first noted by A. B. Migdal~\citep{Migdal:1939}, refers
to an inelastic exit channel in scattering off an atomic nucleus,
where not only the atomic center of mass gets recoil (typically termed
as NR) but also the intrinsic atomic electron state is excited or
ionized. Unlike the elastic exit channel, the Migdal effect generates
other electromagnetic signals in the form of ionized electrons and
photons from atomic de-excitation or recombination, which are more
energetic than NR hence is detectable. This novel DM detection mode
has been applied to several experiments~\citep{Dolan:2017xbu,Kobayashi:2018jky,Akerib:2018hck,Armengaud:2019kfj,Liu:2019kzq,Aprile:2019jmx,Barak:2020fql}
and some of them give the current best limits on DM-nucleus interactions
in the mass range below GeV. With detectors of larger size and longer
data taking time, e.g., xenon-based XENONnT~\citep{Aprile:2020vtw},
LZ~\citep{Akerib:2019fml}, and DARWIN~\citep{Aalbers:2016jon};
argon-based DarkSide~\citep{Aalseth:2015mba}, DEAP~\citep{Amaudruz:2017ibl},
and ArDM~\citep{Calvo:2016hve}; or of lower threshold and better
resolution, \textcolor{black}{e.g., germanium-based EDELWEISS~\citep{Arnaud:2020svb}
and CDMS HVeV~\citep{Agnese:2018col}; silicon-based SENSEI~\citep{Abramoff:2019dfb}
and DAMIC~\citep{Aguilar-Arevalo:2019wdi}}, the Migdal effect will
be a promising probe of hadrophilic LDM.

While detecting the Migdal effect is an experimental challenge, predicting
its count rate is mostly a theoretical one~\citep{Ibe:2017yqa,Bell:2019egg,Essig:2019xkx,Baxter:2019pnz}.
For a relic sub-GeV LDM candidate, its kinetic energy is no bigger
than $\sim\textrm{keV}$, so the scattering process falls in the atomic
scale. Proper understanding of such a sub-keV Migdal effect inevitably
involves atomic physics. The lower the energy, the more pronounced
many-body effects are expected. The complexity of many-body physics
adds a new layer when detector media can no longer be treated as isolated
atoms. This means at certain low-energy levels it will be necessary
to take into account molecular or condensed-matter physics, depending
on a detector's material phase. All these problems are highly non-trivial
but essential.

To address the theory issues related to the Migdal effect, we first
derive a relation that links the low-energy Migdal process to photoabsorption,
which is exact at the long wavelength limit of photon and expected
to work up to $\sim\textrm{keV}$. The relation has two major advantages:
it applies to all kinds of DM detectors in general, and is free of
theoretical uncertainties because only photoabsorption cross section
(experimentally measurable!) is needed for input. To demonstrate the
robustness of this relation, we study the case of xenon by applying
a state-of-the-art atomic approach, the relativistic random phase
approximation (RRPA), whose high precision has been demonstrated in
Ref.~\citep{Chen:2016eab}. Finally we use this relation to predict
the Migdal effects in argon, silicon, and germanium detectors.

\paragraph*{Migdal effect.}

The transition operator and matrix element of a Migdal process has
been derived in several ways, e.g., through a sudden approximation~\citep{Landau_Lifshitz:1991qm},
non-relativistic scattering with Galilean invariance~\citep{Chen:2015pha}
(for hydrogen-like systems), and relativistic scattering with full
Lorentz covariance~\citep{Ibe:2017yqa}. Because of the hierarchy
between atomic, nuclear, electron mass, $m_{A}$, $m_{N}$, $m_{e}$,
and atomic binding energy $E_{B}$: $m_{A}\approx m_{N}\gg m_{e}\gg E_{B}$,
all derivations converge on the resulting Migdal matrix element 
\begin{equation}
M_{FI}=\left\langle F\left|e^{-i\frac{m_{e}}{m_{A}}\vec{q}_{A}\cdot\sum_{i=1}^{Z}\vec{r}_{i}}\right|I\right\rangle \,,\label{eq:Migdal_full}
\end{equation}
where $\vec{q}_{A}$ is the 3-momentum transfer to the atomic system;
$\vec{r}_{i}$ the coordinate of the $i$th electron; $\ket{I}$ and
$\ket{F}$ the atomic initial and final states in the intrinsic frame
(i.e., with the center of mass motion being factored out), respectively.~\footnote{In this work, atomic states are treated relativistically, so the Migdal
operator is a $4\times4$ diagonal matrix.}

First, notice that the summation over all $Z$ electron coordinates
appears in the exponent, so the $n$th-order series expansion contains
$n$-body operators, whose matrix elements are tedious to compute.
By a naïve dimensional analysis: $\left|\vec{q}_{A}\right|\sim m_{\chi}v_{\chi}$
with DM velocity $v_{\chi}\sim10^{-3}$; and $\left\langle \vec{r}_{i}\right\rangle \sim(Z_{i}m_{e}\alpha)^{-1}$
with $Z_{i}$ being the effective charge seen by the $i$th electron
and $\alpha$ the fine structure constant, one can define an atomic-shell-dependent
expansion parameter $\epsilon_{i}=\frac{m_{\chi}}{m_{A}}\frac{v_{\chi}}{\alpha}\frac{1}{Z_{i}}$
. In the case of xenon, $m_{A}\sim120\,\textrm{GeV}$, so $\epsilon_{i}\sim\frac{0.001}{Z_{i}}\frac{m_{\chi}}{\textrm{GeV}}$
guarantees good convergence for sub-GeV DM even with a $Z_{i}=1$
assumption. In reality, current xenon detector thresholds are $\sim\textrm{keV}$,
most Migdal events would be from inner-shell ionizations with $Z_{i}$
surely larger than 1. Therefore, for LDM searches, the Migdal matrix
element can be well-approximated by the leading-order term 
\begin{equation}
M_{FI}^{(1)}=-i\frac{m_{e}}{m_{A}}\vec{q}_{A}\cdot\left\langle F\left|\sum_{i=1}^{Z}\vec{r}_{i}\right|I\right\rangle \equiv-i\frac{m_{e}}{m_{A}}\vec{q}_{A}\cdot\vec{D}_{FI}\,,\label{eq:Migdal_1st}
\end{equation}
where $\vec{D}_{FI}$ is the familiar dipole matrix element. 

The energy deposition by DM in a Migdal process goes into two parts:
one to the atomic center-of-mass kinetic energy (or the NR energy),
$E_{R}$, and the other to the atomic discrete excitation or ionization,
denoted as $E_{r}=E_{F}-E_{I}$ whit $E_{F(I)}$ the eigenenergy of
the state $\ket{F(I)}$. To accommodate different DM detector's approaches
in measuring the combinations of $E_{R}$ and $E_{r}$, it is customary
to cast the cross section in a double differential form 

\begin{equation}
\frac{d\sigma}{dE_{R}dE_{r}}=\frac{m_{e}^{2}}{\mu_{N}^{2}v_{\chi}^{2}}\widetilde{\sigma}_{N}(q_{A})E_{R}\overline{D^{2}}_{FI}\,,\label{eq:ds/dEde}
\end{equation}
where $\mu_{N}=m_{N}m_{\chi}/(m_{N}+m_{\chi})$ is the reduced mass
of the DM-nucleus system, and $\widetilde{\sigma}_{N}$ the DM-nucleus
cross section which depends on $q_{A}=\sqrt{2m_{A}E_{R}}$. The averaged
dipole matrix element squared $\overline{D_{FI}^{2}}$ involves a
summation of all allowed final states and an average of degenerate
initial states.~\footnote{Our choice of wave function normalization is $\braket{E_{I}|E_{I'}}=\delta_{II'}$
for bound states and $\braket{E_{F}|E_{F'}}=\delta(E_{F}-E_{F'})$
for continuum states.} 

\paragraph*{Photoabsorption and its relation to the Migdal effect.}

As seen from Eq.~\ref{eq:ds/dEde}, the quantity $\overline{D^{2}}_{FI}$
is the critical piece of information that requires many-body calculations.
Therefore, it is desirable to ask whether it can be extracted directly
from experiments. Fortunately, photoabsorption provides the answer.

The full transition matrix element for photoabsorption takes the form
\begin{equation}
P_{FI}=\hat{\varepsilon}\cdot\left\langle F\left|\sum_{i=1}^{Z}e^{i\vec{k}\cdot\vec{r}_{i}}\vec{\alpha}_{i}\right|I\right\rangle \equiv\hat{\varepsilon}\cdot\vec{O}_{FI},\label{eq:me_photo}
\end{equation}
where $\vec{k}$ and $\hat{\varepsilon}$ are the photon momentum
and polarization vectors, and $\vec{\alpha}_{i}$ the $4\times4$
spatial Dirac matrix. By energy conservation, the photon energy $\omega=|\vec{k}|=E_{r}$
(the atomic recoil is negligible as $\omega\ll m_{A}$). The total
cross section of photoabsorption is 
\begin{equation}
\sigma_{\gamma}(E_{r})=\frac{4\pi^{2}\alpha}{E_{r}}\overline{P_{FI}^{2}}\,,\label{eq:sigma_gamma}
\end{equation}
where the same summation and average of states being applied to $P_{FI}$
as in the case to $D_{FI}$. 

The standard procedure of calculating $\sigma_{\gamma}(E_{r})$ starts
by a multipole expansion of $P_{FI}$ that yields the transverse electric,
$T_{J}^{el}(kr)$, and transverse magnetic, $T_{J}^{mag}(kr)$, multipoles
with $J=1,2,\ldots$ denoting the spherical multipolarity. At the
long wavelength (LW) limit, i.e., $k\left\langle r\right\rangle \ll1$
so that $e^{i\vec{k}\cdot\vec{r}}\rightarrow1$, the transition matrix
element is simplified to 
\begin{equation}
\vec{O}_{FI}^{(E1)}=\left\langle F\left|\sum_{i=1}^{Z}\vec{\alpha}_{i}\right|I\right\rangle \equiv\vec{D}_{FI}^{(\textrm{V})}=iE_{r}\vec{D}_{FI}.\label{eq:me_photo_LW}
\end{equation}
This is the electric dipole ($E1$) approximation for photoabsorption,
as the resulting atomic operator is a parity-odd dipole, either as
$\vec{\alpha}_{i}$, usually termed as the velocity form (denoted
by the superscript ``V''), or $\vec{r}_{i}$, the length form. The
equivalence of these two operators is established by the commutation
relation $-i[\vec{r}_{i},H]=\vec{\alpha}_{i}$. As a result, $\sigma_{\gamma}(E_{r})$
can be approximated by 

\begin{equation}
\sigma_{\gamma}(E_{r})\xrightarrow[E1\textrm{ approx.}]{}4\pi^{2}\alpha E_{r}\overline{D^{2}}_{FI}\,,\label{eq:sigma_gamma_E1}
\end{equation}
and the Migdal differential cross section can be cast into 

\begin{equation}
\frac{d\sigma^{(\textrm{MPA})}}{dE_{R}dE_{r}}=\frac{m_{e}^{2}}{\mu_{N}^{2}v_{\chi}^{2}}\tilde{\sigma}_{N}(q_{A})\frac{E_{R}}{E_{r}}\frac{\sigma_{\gamma}(E_{r})}{4\pi^{2}\alpha}\,,\label{eq:MPA}
\end{equation}
which we call the ``Migdal-photoabsorption'' (MPA) relation. This
relation is powerful: it tells that as long as the photoabsorption
cross section can be measured in a detector, the corresponding Migdal
effect can be figured accordingly, and there is no need for theory
input. In contrast, the relation proposed in Ref.~\citep{Essig:2019xkx}
involves DM-electron scattering, which is yet to be detected.

While the derivation of the $E1$ approximation is straightforward,
there are two points of particular importance from a theory viewpoint.
First, at the LW limit, all the so-called retardation effects from
$e^{i\vec{k}\cdot\vec{r}}-1$ are ignored. They are grouped into higher-rank
spherical multipoles or higher-order corrections (in powers of $k^{2}r^{2}$)
to spherical multipoles of a given rank. 

Second, the equivalence between the dipole operators in the velocity
and length forms, manifested here by a simple commutation relation,
has a deeper connection to the gauge invariance of electromagnetism.
As first noted by Siegert~\citep{Siegert:1939zz}, the transverse
electric multipole operators can be related by current conservation
to the charge multipole operators at the LW limit. However, the equivalence
at the matrix-element level has an additional requirement that the
wave functions are energy eigenstates, i.e., $H\ket{F(I)}=E_{F(I)}\ket{F(I)}$.
For most many-body calculations that only approximate the true eigenstates,
the breaking of gauge invariance, e.g., $\vec{D}_{FI}^{(\textrm{V})}\neq\vec{D}_{FI}$,
is quite commonly seen. Therefore, adopting many-body approaches that
preserve gauge invariance, such as (R)RPA~\citep{Lin:1977dl},~\footnote{In the same paper~\citep{Lin:1977dl}, the most commonly-used Hartree-Fock
method is shown to violate gauge invariance.} is preferred. Conversely, the degree of broken gauge invariance can
serve as a robustness test of a many-body calculation. In atomic physics,
this is usually done with two different forms of $T_{J}^{el}$, one
in the Coulomb gauge and the other the ``length gauge''~\citep{Grant_1974}. 

\paragraph*{Case study of xenon. }

\begin{figure}
\begin{center}\includegraphics[width=1\columnwidth]{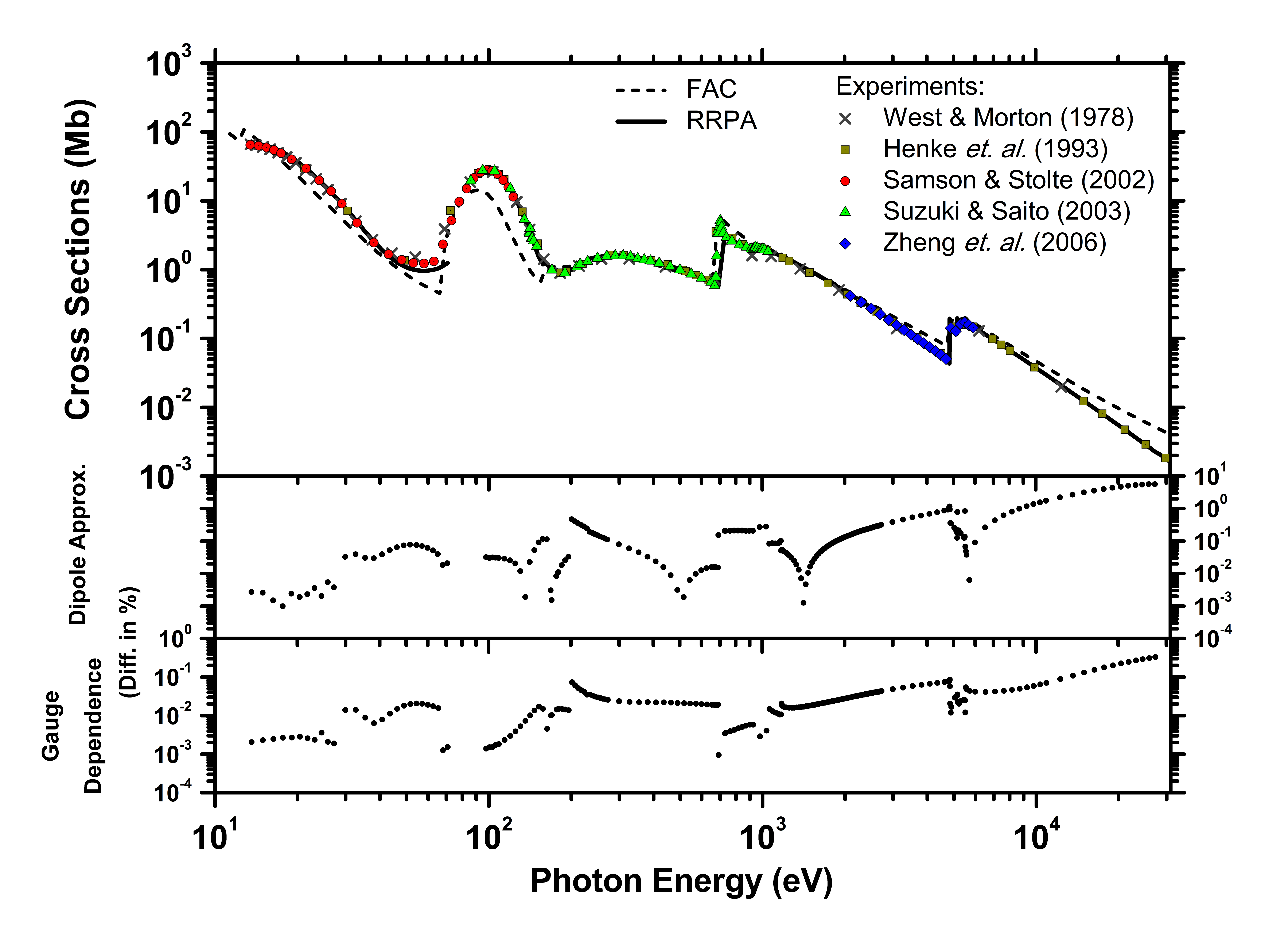}

\end{center}

\caption{(Top panel) Xenon photoabsorption cross section from experiments and
atomic calculations of RRPA with the length-gauge operators and the
FAC code. Also shown are the percentage differences by using the $E1$
approximation (central panel) and from using the Coulomb gauge operators
(bottom panel) in the same RRPA routine.~\label{fig:photoabs}}
\end{figure}

\begin{figure*}
\begin{center}%
\begin{tabular}{cc}
\includegraphics[width=0.5\textwidth]{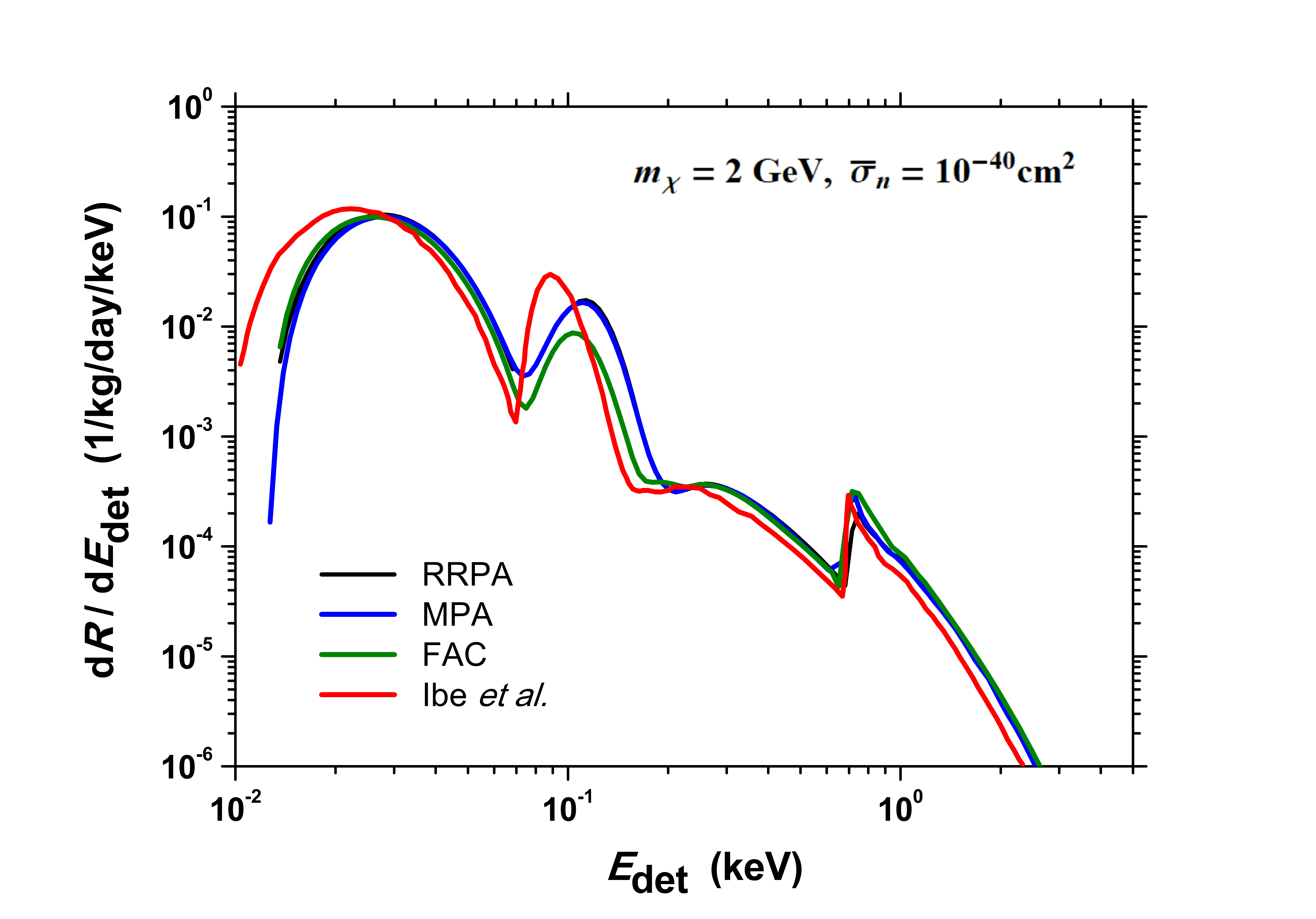} & \includegraphics[width=0.5\textwidth]{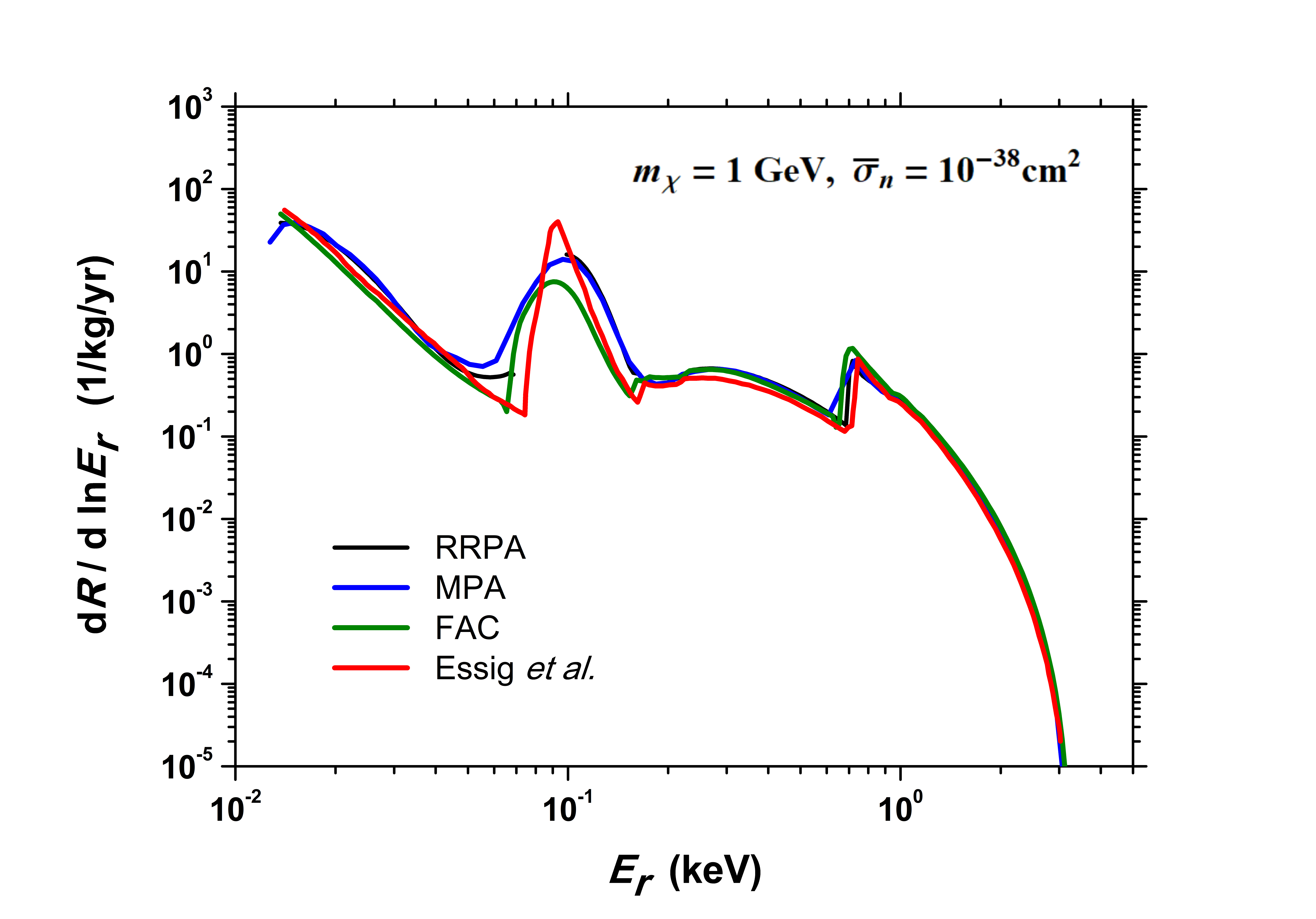}\tabularnewline
\end{tabular}

\end{center}

\caption{Differential count rates $\frac{dR}{dE_{\det}}$ (left) and $\frac{dR}{d\ln E_{r}}$
(right) of the Migdal effect in xenon detectors by the spin-independent,
isoscalar, DM-nucleon contact interaction (with cross section $\bar{\sigma}_{n}$
) predicted by (i) RRPA, (ii) MPA relation, (iii) FAC, and (iv) Ref.~\citep{Ibe:2017yqa}
(left) and Ref.~\citep{Essig:2019xkx} (right).~\label{fig:migdal}}
\end{figure*}

In Fig.~\ref{fig:photoabs}, the experimental data for xenon photoabsorption,
compiled from Refs.~\citep{West:1978xe,Henke:1993gd,Samson:2002xe,Suzuki:2003xe,Zheng:2006xe},
are compared with several theoretical calculations. The agreement
between the RRPA curve, taken from Ref.~\citep{Chen:2016eab} using
operators in the length gauge, shows that our atomic approach can
handle many-body excited states properly. In this work, we carry out
two additional calculations: one with the $E1$ approximation, i.e.,
Eq.~(\ref{eq:sigma_gamma_E1}), and the other with operators in the
Coulomb gauge. As shown in the central panel, the $E1$ approximation
works very well up to $1\,\textrm{keV}$ with all higher-order corrections
still kept at a level below $1\%$. This justifies the basic assumption
underlying the MPA relation: the averaged dipole matrix element squared
$\overline{D_{FI}^{2}}$ in the sub-keV Migdal process can be reliably
extracted from photoabsorption measurements. In the bottom panel,
the nontrivial property of gauge invariance in many-body calculations
is clearly shown to be preserved by the RRPA approach.

The FAC results are obtained by running a built-in module in the ``Flexible
Atomic Code'' package that calculates photoionization cross sections
directly~\citep{Gu:2008fac}. As the comparison shows, the FAC code
does a reasonably good job for $E_{r}\gtrsim200\,\textrm{eV}$ in
general, but at lower energies, it does not perform well, and errors
at some points are quite large. By construct, the FAC package is mainly
designed for highly-ionized atoms and built with focus on efficiency
instead of accuracy. Therefore, its many-body approach is solving
a prescribed form of averaged one-body potential (in this sense, not
an \textit{ab inito} approach) self-consistently, and leads to a picture
that all electrons act like independent particles, which is unrealistic
at low energies. Another noteworthy point is for $E_{r}\gtrsim1\,\textrm{keV}$,
FAC has the tendency to over-predict when $E_{r}$ gets away from
edge energies. This is because the transition operators being adopted,
except $T_{1}^{mag}$, are non-relativistic, and contributions from
all sub-leading orders in $k^{2}r^{2}$ are missing. 

For prediction of Migdal count rates, we follow the same procedures
as in Ref.~\citep{Ibe:2017yqa} and \citep{Essig:2019xkx}. In the
former case, the observable energy $E_{\textrm{det}}$ is a sum of
$E_{r}$ and $q_{\textrm{nr}}E_{R}$ with $q_{\textrm{nr}}$ the NR
quenching factor, and the differential count rate is 
\begin{align}
\frac{dR}{dE_{\textrm{det}}}= & n_{\chi}N_{T}\int dE_{R}\int dE_{r}\delta(E_{\textrm{det}}-q_{\textrm{nr}}E_{R}-E_{r})\nonumber \\
 & \times\tilde{\sigma}_{N}(q_{A})E_{R}\overline{D^{2}}_{FI}\eta(v_{\textrm{min}})\,.\label{eq:dRdEdet}
\end{align}
For the latter case, the observable energy is $E_{r}$ and the differential
count rate is simply 
\begin{equation}
\frac{dR}{dE_{r}}=n_{\chi}N_{T}\frac{m_{e}^{2}}{\mu_{N}^{2}}\overline{D^{2}}_{FI}\int dE_{R}\,\tilde{\sigma}_{N}(q_{A})E_{R}\eta(v_{\textrm{min}})\,,\label{eq:dRdEr}
\end{equation}
where $n_{\chi}$ is the local DM number density, $N_{T}$ the number
of target atoms, and the $\eta$ function results from the $1/v_{\chi}$
factor averaged with the DM velocity spectrum~\citep{Lewin:1995rx},
and depends on the minimum DM velocity $v_{\min}=(m_{N}E_{R}+\mu_{N}E_{r})/(\mu_{N}\sqrt{2m_{N}E_{R}})$
that guarantees energy deposition of $E_{R}+E_{r}$ is possible. 

In Fig.~\ref{fig:migdal}, we plot four sets of count rate predictions
assuming a contact, spin-independent, isoscalar DM-nucleus interaction
so that $\tilde{\sigma}_{N}=A^{2}\mu_{N}^{2}\mu_{n}^{-2}\bar{\sigma}_{n}$
with $\mu_{n}$ being the DM-nucleon reduced mass and $\bar{\sigma}_{n}$
the DM-nucleon cross section: The black line is by direct computation
of Eq.~(\ref{eq:ds/dEde}) using RRPA; the blue (green) line is obtained
by the MPA relation with $\sigma_{\gamma}(E_{r})$ taken from data
(FAC). The nice agreement between RRPA and MPA is not only a justification
to the applicability of the MPA relation, but also a theory-experiment
double confirmation of the results. The difference between our results
from the ones of Refs.~\citep{Ibe:2017yqa,Essig:2019xkx} are most
likely originated from different atomic approaches. Generally speaking,
mean field methods can give good results of ground state properties
and wave functions, but their applicability to excited states can
be problematic. \textcolor{black}{Note that Ref.~\citep{Ibe:2017yqa}
used the FAC package differently from what we do. The authors adopted
the picture that the atomic excited states are purely 1-particle-1-hole
excitations (because the leading-order Migdal operator is one-body)
from the ground state. This independent particle picture ignores not
only the residual two-body correlation, but also the fact that atomic
mean field varies with electronic configuration. The FAC package takes
into account the latter aspect by diagonalizing the atomic Hamiltonian
in the model space of a given problem. The resulting wave functions
are configuration-mixed, i.e., not in form of a single Slater determinant.
}This explains why the FAC results are closer to our RRPA results
than \textcolor{black}{Ref.~\citep{Ibe:2017yqa}}.

\paragraph*{Applications to argon, silicon, and germanium detectors.}

\begin{figure}
\includegraphics[width=1\columnwidth]{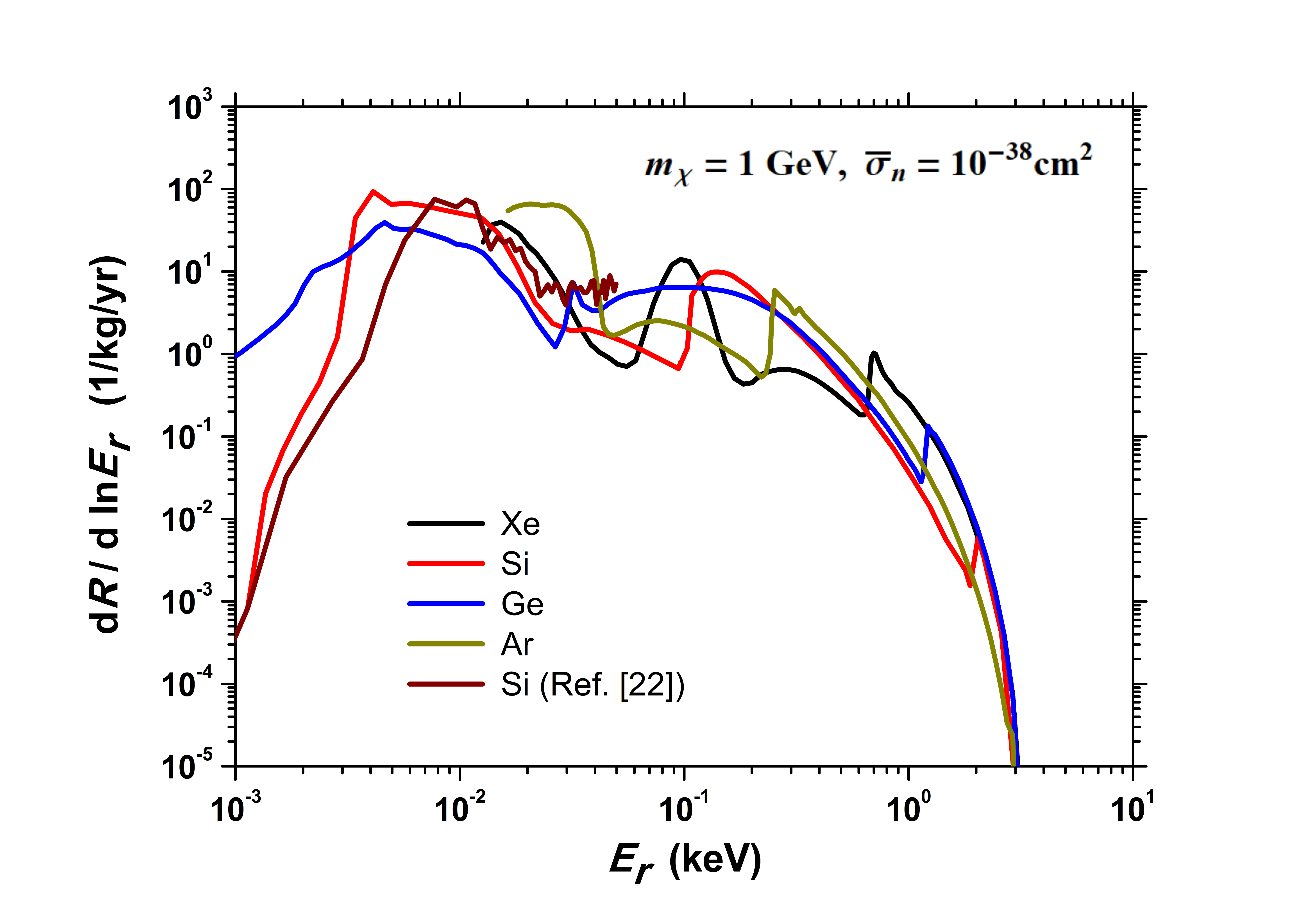}

\caption{Differential count rates $\frac{dR}{d\ln E_{r}}$ for xenon, argon,
semiconductor silicon and germanium detectors. \label{fig:comp}}
\end{figure}

Using the MPA relation, we combine in Fig.~\ref{fig:comp} the predicted
count rates for xenon, argon, semiconductor silicon and germanium
detectors. For $E_{r}\ge10\,\textrm{eV}$, the measured photoabsorption
data are taken from Ref.~\citep{Henke:1993gd} along with semi-empirical
fitting; for semiconductor silicon and germanium at room temperature
and $1\,\textrm{eV}<E_{r}<10\,\textrm{eV}$, they are from Ref.~\citep{Sze:1981sem}
and \citep{Philipp:1959ge}, respectively. The silicon prediction
roughly agrees with the one of Ref.~\citep{Essig:2019xkx}, which
is based on a detailed condensed matter calculation. The difference
in the $1-5\,\textrm{eV}$ region could be from the many-body computation
or because the photoabsorption data being used do not exactly apply
to the case of the SENSEI detector. The argon prediction should be
robust, unlike the large theoretical uncertainties assigned in Ref.~\citep{GrillidiCortona:2020owp}
based on the atomic calculation of Ref.~\citep{Ibe:2017yqa}.

More impressively, the MPA relation provides a reliable germanium
prediction in the energy range of $1-80\,\textrm{eV}$. According
to our previous study of atomic germanium~\citep{Chen:2013lba,Chen:2014ypv},
the MCRRPA method (the RRPA with multiconfiguration required for open-shell
atoms), though sophisticated enough, does not work satisfactorily
in this energy because the crystal effects. Now, it is no longer an
obstacle, and the experimental analyses, such as being done by EDELWEISS~\citep{Armengaud:2019kfj}
and CDEX~\citep{Liu:2019kzq}, can include these shells, which have
dominant contributions for $E_{r}<100\,\textrm{eV}$. Another significant
feature is that the germanium detector can be very sensitive to the
Migdal effect at extremely low energy, because of its specially large
photoabsorption coefficient in $E_{r}=1-3\,\textrm{eV}$. Both CDMS
HVeV~\citep{Agnese:2018col} and EDELWEISS~\citep{Arnaud:2020svb}
recently demonstrate their extremely-low-threshold capability at $1\,\textrm{eV}$.
According to the plot, the Migdal count rate can be 3-order-of-magnitude
bigger than in a silicon detector, assuming equal exposure mass-time.

It should also be pointed out that different detectors complement
one another, thanks to the rich atomic structure. At each photoabsorption
peak, the cross section and the resulting Migdal rate can receive
an substantial boost. This not only enhances the sensitivity but also
provides smoking gun signatures.

\paragraph*{Conclusion.}

The Migdal effect has been a powerful search mode for hadrophilic,
sub-GeV light dark matter. With modern technologies pushing detector
thresholds lower and lower, our range of light dark matter searches
will expand further. The ``Migdal-photoabsorption'' relation we
derived in this work is based on general principles and only requires
photoabsorption measurements as input. It thus provides predictions
that have no uncertainties from many-body calculations. Though we
only consider xenon, argon, semiconductor silicon and germanium detectors
as examples, it certainly can be applied to other novel low-threshold
detectors.
\begin{acknowledgments}
This work is supported in part under Grant Nos. 108-2112-M-259-003
(CPL) , 108-2112-M-002-003-MY3 (JWC) from the Ministry of Science
and Technology, 2019-20/ECP-2 from the National Center of Theoretical
Sciences, and Kenda Foundation (JWC) of Taiwan; and the Canada First
Research Excellence Fund through the Arthur B. McDonald Canadian Astroparticle
Physics Research Institute (CPW).
\end{acknowledgments}

\bibliographystyle{apsrev4-1}
\bibliography{../Migdal}

\begin{thebibliography}{41}%
\makeatletter
\providecommand \@ifxundefined [1]{%
 \@ifx{#1\undefined}
}%
\providecommand \@ifnum [1]{%
 \ifnum #1\expandafter \@firstoftwo
 \else \expandafter \@secondoftwo
 \fi
}%
\providecommand \@ifx [1]{%
 \ifx #1\expandafter \@firstoftwo
 \else \expandafter \@secondoftwo
 \fi
}%
\providecommand \natexlab [1]{#1}%
\providecommand \enquote  [1]{``#1''}%
\providecommand \bibnamefont  [1]{#1}%
\providecommand \bibfnamefont [1]{#1}%
\providecommand \citenamefont [1]{#1}%
\providecommand \href@noop [0]{\@secondoftwo}%
\providecommand \href [0]{\begingroup \@sanitize@url \@href}%
\providecommand \@href[1]{\@@startlink{#1}\@@href}%
\providecommand \@@href[1]{\endgroup#1\@@endlink}%
\providecommand \@sanitize@url [0]{\catcode `\\12\catcode `\$12\catcode
  `\&12\catcode `\#12\catcode `\^12\catcode `\_12\catcode `\%12\relax}%
\providecommand \@@startlink[1]{}%
\providecommand \@@endlink[0]{}%
\providecommand \url  [0]{\begingroup\@sanitize@url \@url }%
\providecommand \@url [1]{\endgroup\@href {#1}{\urlprefix }}%
\providecommand \urlprefix  [0]{URL }%
\providecommand \Eprint [0]{\href }%
\providecommand \doibase [0]{http://dx.doi.org/}%
\providecommand \selectlanguage [0]{\@gobble}%
\providecommand \bibinfo  [0]{\@secondoftwo}%
\providecommand \bibfield  [0]{\@secondoftwo}%
\providecommand \translation [1]{[#1]}%
\providecommand \BibitemOpen [0]{}%
\providecommand \bibitemStop [0]{}%
\providecommand \bibitemNoStop [0]{.\EOS\space}%
\providecommand \EOS [0]{\spacefactor3000\relax}%
\providecommand \BibitemShut  [1]{\csname bibitem#1\endcsname}%
\let\auto@bib@innerbib\@empty
\bibitem [{\citenamefont {Zyla}\ \emph {et~al.}(2020)\citenamefont {Zyla} \emph
  {et~al.}}]{Zyla:2020pdg}%
  \BibitemOpen
  \bibfield  {author} {\bibinfo {author} {\bibfnamefont {P.~A.}\ \bibnamefont
  {Zyla}} \emph {et~al.} (\bibinfo {collaboration} {Particle Data Group}),\
  }\href@noop {} {\bibfield  {journal} {\bibinfo  {journal} {Prog. Theor. Exp.
  Phys.}\ ,\ \bibinfo {pages} {083C01}} (\bibinfo {year} {2020})},\ \bibinfo
  {note} {to be published}\BibitemShut {NoStop}%
\bibitem [{\citenamefont {Ibe}\ \emph {et~al.}(2018)\citenamefont {Ibe},
  \citenamefont {Nakano}, \citenamefont {Shoji},\ and\ \citenamefont
  {Suzuki}}]{Ibe:2017yqa}%
  \BibitemOpen
  \bibfield  {author} {\bibinfo {author} {\bibfnamefont {M.}~\bibnamefont
  {Ibe}}, \bibinfo {author} {\bibfnamefont {W.}~\bibnamefont {Nakano}},
  \bibinfo {author} {\bibfnamefont {Y.}~\bibnamefont {Shoji}}, \ and\ \bibinfo
  {author} {\bibfnamefont {K.}~\bibnamefont {Suzuki}},\ }\href {\doibase
  10.1007/JHEP03(2018)194} {\bibfield  {journal} {\bibinfo  {journal} {JHEP}\
  }\textbf {\bibinfo {volume} {03}},\ \bibinfo {pages} {194} (\bibinfo {year}
  {2018})},\ \Eprint {http://arxiv.org/abs/1707.07258} {arXiv:1707.07258
  [hep-ph]} \BibitemShut {NoStop}%
\bibitem [{\citenamefont {Migdal}(1939)}]{Migdal:1939}%
  \BibitemOpen
  \bibfield  {author} {\bibinfo {author} {\bibfnamefont {A.~B.}\ \bibnamefont
  {Migdal}},\ }\href@noop {} {\bibfield  {journal} {\bibinfo  {journal}
  {ZhETF}\ }\textbf {\bibinfo {volume} {9}},\ \bibinfo {pages} {1163} (\bibinfo
  {year} {1939})}\BibitemShut {NoStop}%
\bibitem [{\citenamefont {Dolan}\ \emph {et~al.}(2018)\citenamefont {Dolan},
  \citenamefont {Kahlhoefer},\ and\ \citenamefont {McCabe}}]{Dolan:2017xbu}%
  \BibitemOpen
  \bibfield  {author} {\bibinfo {author} {\bibfnamefont {M.~J.}\ \bibnamefont
  {Dolan}}, \bibinfo {author} {\bibfnamefont {F.}~\bibnamefont {Kahlhoefer}}, \
  and\ \bibinfo {author} {\bibfnamefont {C.}~\bibnamefont {McCabe}},\ }\href
  {\doibase 10.1103/PhysRevLett.121.101801} {\bibfield  {journal} {\bibinfo
  {journal} {Phys. Rev. Lett.}\ }\textbf {\bibinfo {volume} {121}},\ \bibinfo
  {pages} {101801} (\bibinfo {year} {2018})},\ \Eprint
  {http://arxiv.org/abs/1711.09906} {arXiv:1711.09906 [hep-ph]} \BibitemShut
  {NoStop}%
\bibitem [{\citenamefont {Kobayashi}\ \emph {et~al.}(2019)\citenamefont
  {Kobayashi} \emph {et~al.}}]{Kobayashi:2018jky}%
  \BibitemOpen
  \bibfield  {author} {\bibinfo {author} {\bibfnamefont {M.}~\bibnamefont
  {Kobayashi}} \emph {et~al.} (\bibinfo {collaboration} {XMASS}),\ }\href
  {\doibase 10.1016/j.physletb.2019.06.022} {\bibfield  {journal} {\bibinfo
  {journal} {Phys. Lett. B}\ }\textbf {\bibinfo {volume} {795}},\ \bibinfo
  {pages} {308} (\bibinfo {year} {2019})},\ \Eprint
  {http://arxiv.org/abs/1808.06177} {arXiv:1808.06177 [astro-ph.CO]}
  \BibitemShut {NoStop}%
\bibitem [{\citenamefont {Akerib}\ \emph {et~al.}(2019)\citenamefont {Akerib}
  \emph {et~al.}}]{Akerib:2018hck}%
  \BibitemOpen
  \bibfield  {author} {\bibinfo {author} {\bibfnamefont {D.}~\bibnamefont
  {Akerib}} \emph {et~al.} (\bibinfo {collaboration} {LUX}),\ }\href {\doibase
  10.1103/PhysRevLett.122.131301} {\bibfield  {journal} {\bibinfo  {journal}
  {Phys. Rev. Lett.}\ }\textbf {\bibinfo {volume} {122}},\ \bibinfo {pages}
  {131301} (\bibinfo {year} {2019})},\ \Eprint
  {http://arxiv.org/abs/1811.11241} {arXiv:1811.11241 [astro-ph.CO]}
  \BibitemShut {NoStop}%
\bibitem [{\citenamefont {Armengaud}\ \emph {et~al.}(2019)\citenamefont
  {Armengaud} \emph {et~al.}}]{Armengaud:2019kfj}%
  \BibitemOpen
  \bibfield  {author} {\bibinfo {author} {\bibfnamefont {E.}~\bibnamefont
  {Armengaud}} \emph {et~al.} (\bibinfo {collaboration} {EDELWEISS}),\ }\href
  {\doibase 10.1103/PhysRevD.99.082003} {\bibfield  {journal} {\bibinfo
  {journal} {Phys. Rev. D}\ }\textbf {\bibinfo {volume} {99}},\ \bibinfo
  {pages} {082003} (\bibinfo {year} {2019})},\ \Eprint
  {http://arxiv.org/abs/1901.03588} {arXiv:1901.03588 [astro-ph.GA]}
  \BibitemShut {NoStop}%
\bibitem [{\citenamefont {Liu}\ \emph {et~al.}(2019)\citenamefont {Liu} \emph
  {et~al.}}]{Liu:2019kzq}%
  \BibitemOpen
  \bibfield  {author} {\bibinfo {author} {\bibfnamefont {Z.}~\bibnamefont
  {Liu}} \emph {et~al.} (\bibinfo {collaboration} {CDEX}),\ }\href {\doibase
  10.1103/PhysRevLett.123.161301} {\bibfield  {journal} {\bibinfo  {journal}
  {Phys. Rev. Lett.}\ }\textbf {\bibinfo {volume} {123}},\ \bibinfo {pages}
  {161301} (\bibinfo {year} {2019})},\ \Eprint
  {http://arxiv.org/abs/1905.00354} {arXiv:1905.00354 [hep-ex]} \BibitemShut
  {NoStop}%
\bibitem [{\citenamefont {Aprile}\ \emph {et~al.}(2019)\citenamefont {Aprile}
  \emph {et~al.}}]{Aprile:2019jmx}%
  \BibitemOpen
  \bibfield  {author} {\bibinfo {author} {\bibfnamefont {E.}~\bibnamefont
  {Aprile}} \emph {et~al.} (\bibinfo {collaboration} {XENON}),\ }\href
  {\doibase 10.1103/PhysRevLett.123.241803} {\bibfield  {journal} {\bibinfo
  {journal} {Phys. Rev. Lett.}\ }\textbf {\bibinfo {volume} {123}},\ \bibinfo
  {pages} {241803} (\bibinfo {year} {2019})},\ \Eprint
  {http://arxiv.org/abs/1907.12771} {arXiv:1907.12771 [hep-ex]} \BibitemShut
  {NoStop}%
\bibitem [{\citenamefont {Barak}\ \emph {et~al.}(2020)\citenamefont {Barak}
  \emph {et~al.}}]{Barak:2020fql}%
  \BibitemOpen
  \bibfield  {author} {\bibinfo {author} {\bibfnamefont {L.}~\bibnamefont
  {Barak}} \emph {et~al.} (\bibinfo {collaboration} {SENSEI}),\ }\href@noop {}
  {\  (\bibinfo {year} {2020})},\ \Eprint {http://arxiv.org/abs/2004.11378}
  {arXiv:2004.11378 [astro-ph.CO]} \BibitemShut {NoStop}%
\bibitem [{\citenamefont {Aprile}\ \emph {et~al.}(2020)\citenamefont {Aprile}
  \emph {et~al.}}]{Aprile:2020vtw}%
  \BibitemOpen
  \bibfield  {author} {\bibinfo {author} {\bibfnamefont {E.}~\bibnamefont
  {Aprile}} \emph {et~al.} (\bibinfo {collaboration} {XENON}),\ }\href@noop {}
  {\  (\bibinfo {year} {2020})},\ \Eprint {http://arxiv.org/abs/2007.08796}
  {arXiv:2007.08796 [physics.ins-det]} \BibitemShut {NoStop}%
\bibitem [{\citenamefont {Akerib}\ \emph {et~al.}(2020)\citenamefont {Akerib}
  \emph {et~al.}}]{Akerib:2019fml}%
  \BibitemOpen
  \bibfield  {author} {\bibinfo {author} {\bibfnamefont {D.}~\bibnamefont
  {Akerib}} \emph {et~al.} (\bibinfo {collaboration} {LZ}),\ }\href {\doibase
  10.1016/j.nima.2019.163047} {\bibfield  {journal} {\bibinfo  {journal} {Nucl.
  Instrum. Meth. A}\ }\textbf {\bibinfo {volume} {953}},\ \bibinfo {pages}
  {163047} (\bibinfo {year} {2020})},\ \Eprint
  {http://arxiv.org/abs/1910.09124} {arXiv:1910.09124 [physics.ins-det]}
  \BibitemShut {NoStop}%
\bibitem [{\citenamefont {Aalbers}\ \emph {et~al.}(2016)\citenamefont {Aalbers}
  \emph {et~al.}}]{Aalbers:2016jon}%
  \BibitemOpen
  \bibfield  {author} {\bibinfo {author} {\bibfnamefont {J.}~\bibnamefont
  {Aalbers}} \emph {et~al.} (\bibinfo {collaboration} {DARWIN}),\ }\href
  {\doibase 10.1088/1475-7516/2016/11/017} {\bibfield  {journal} {\bibinfo
  {journal} {JCAP}\ }\textbf {\bibinfo {volume} {11}},\ \bibinfo {pages} {017}
  (\bibinfo {year} {2016})},\ \Eprint {http://arxiv.org/abs/1606.07001}
  {arXiv:1606.07001 [astro-ph.IM]} \BibitemShut {NoStop}%
\bibitem [{\citenamefont {Aalseth}\ \emph {et~al.}(2015)\citenamefont {Aalseth}
  \emph {et~al.}}]{Aalseth:2015mba}%
  \BibitemOpen
  \bibfield  {author} {\bibinfo {author} {\bibfnamefont {C.}~\bibnamefont
  {Aalseth}} \emph {et~al.},\ }\href {\doibase 10.1155/2015/541362} {\bibfield
  {journal} {\bibinfo  {journal} {Adv. High Energy Phys.}\ }\textbf {\bibinfo
  {volume} {2015}},\ \bibinfo {pages} {541362} (\bibinfo {year}
  {2015})}\BibitemShut {NoStop}%
\bibitem [{\citenamefont {Amaudruz}\ \emph {et~al.}(2019)\citenamefont
  {Amaudruz} \emph {et~al.}}]{Amaudruz:2017ibl}%
  \BibitemOpen
  \bibfield  {author} {\bibinfo {author} {\bibfnamefont {P.-A.}\ \bibnamefont
  {Amaudruz}} \emph {et~al.} (\bibinfo {collaboration} {DEAP-3600}),\ }\href
  {\doibase 10.1016/j.astropartphys.2018.09.006} {\bibfield  {journal}
  {\bibinfo  {journal} {Astropart. Phys.}\ }\textbf {\bibinfo {volume} {108}},\
  \bibinfo {pages} {1} (\bibinfo {year} {2019})},\ \Eprint
  {http://arxiv.org/abs/1712.01982} {arXiv:1712.01982 [astro-ph.IM]}
  \BibitemShut {NoStop}%
\bibitem [{\citenamefont {Calvo}\ \emph {et~al.}(2017)\citenamefont {Calvo}
  \emph {et~al.}}]{Calvo:2016hve}%
  \BibitemOpen
  \bibfield  {author} {\bibinfo {author} {\bibfnamefont {J.}~\bibnamefont
  {Calvo}} \emph {et~al.} (\bibinfo {collaboration} {ArDM}),\ }\href {\doibase
  10.1088/1475-7516/2017/03/003} {\bibfield  {journal} {\bibinfo  {journal}
  {JCAP}\ }\textbf {\bibinfo {volume} {03}},\ \bibinfo {pages} {003} (\bibinfo
  {year} {2017})},\ \Eprint {http://arxiv.org/abs/1612.06375} {arXiv:1612.06375
  [physics.ins-det]} \BibitemShut {NoStop}%
\bibitem [{\citenamefont {Arnaud}\ \emph {et~al.}(2020)\citenamefont {Arnaud}
  \emph {et~al.}}]{Arnaud:2020svb}%
  \BibitemOpen
  \bibfield  {author} {\bibinfo {author} {\bibfnamefont {Q.}~\bibnamefont
  {Arnaud}} \emph {et~al.} (\bibinfo {collaboration} {EDELWEISS}),\ }\href@noop
  {} {\  (\bibinfo {year} {2020})},\ \Eprint {http://arxiv.org/abs/2003.01046}
  {arXiv:2003.01046 [astro-ph.GA]} \BibitemShut {NoStop}%
\bibitem [{\citenamefont {Agnese}\ \emph {et~al.}(2018)\citenamefont {Agnese}
  \emph {et~al.}}]{Agnese:2018col}%
  \BibitemOpen
  \bibfield  {author} {\bibinfo {author} {\bibfnamefont {R.}~\bibnamefont
  {Agnese}} \emph {et~al.} (\bibinfo {collaboration} {SuperCDMS}),\ }\href
  {\doibase 10.1103/PhysRevLett.121.051301} {\bibfield  {journal} {\bibinfo
  {journal} {Phys. Rev. Lett.}\ }\textbf {\bibinfo {volume} {121}},\ \bibinfo
  {pages} {051301} (\bibinfo {year} {2018})},\ \bibinfo {note} {[Erratum:
  Phys.Rev.Lett. 122, 069901 (2019)]},\ \Eprint
  {http://arxiv.org/abs/1804.10697} {arXiv:1804.10697 [hep-ex]} \BibitemShut
  {NoStop}%
\bibitem [{\citenamefont {Abramoff}\ \emph {et~al.}(2019)\citenamefont
  {Abramoff} \emph {et~al.}}]{Abramoff:2019dfb}%
  \BibitemOpen
  \bibfield  {author} {\bibinfo {author} {\bibfnamefont {O.}~\bibnamefont
  {Abramoff}} \emph {et~al.} (\bibinfo {collaboration} {SENSEI}),\ }\href
  {\doibase 10.1103/PhysRevLett.122.161801} {\bibfield  {journal} {\bibinfo
  {journal} {Phys. Rev. Lett.}\ }\textbf {\bibinfo {volume} {122}},\ \bibinfo
  {pages} {161801} (\bibinfo {year} {2019})},\ \Eprint
  {http://arxiv.org/abs/1901.10478} {arXiv:1901.10478 [hep-ex]} \BibitemShut
  {NoStop}%
\bibitem [{\citenamefont {Aguilar-Arevalo}\ \emph {et~al.}(2019)\citenamefont
  {Aguilar-Arevalo} \emph {et~al.}}]{Aguilar-Arevalo:2019wdi}%
  \BibitemOpen
  \bibfield  {author} {\bibinfo {author} {\bibfnamefont {A.}~\bibnamefont
  {Aguilar-Arevalo}} \emph {et~al.} (\bibinfo {collaboration} {DAMIC}),\ }\href
  {\doibase 10.1103/PhysRevLett.123.181802} {\bibfield  {journal} {\bibinfo
  {journal} {Phys. Rev. Lett.}\ }\textbf {\bibinfo {volume} {123}},\ \bibinfo
  {pages} {181802} (\bibinfo {year} {2019})},\ \Eprint
  {http://arxiv.org/abs/1907.12628} {arXiv:1907.12628 [astro-ph.CO]}
  \BibitemShut {NoStop}%
\bibitem [{\citenamefont {Bell}\ \emph {et~al.}(2020)\citenamefont {Bell},
  \citenamefont {Dent}, \citenamefont {Newstead}, \citenamefont {Sabharwal},\
  and\ \citenamefont {Weiler}}]{Bell:2019egg}%
  \BibitemOpen
  \bibfield  {author} {\bibinfo {author} {\bibfnamefont {N.~F.}\ \bibnamefont
  {Bell}}, \bibinfo {author} {\bibfnamefont {J.~B.}\ \bibnamefont {Dent}},
  \bibinfo {author} {\bibfnamefont {J.~L.}\ \bibnamefont {Newstead}}, \bibinfo
  {author} {\bibfnamefont {S.}~\bibnamefont {Sabharwal}}, \ and\ \bibinfo
  {author} {\bibfnamefont {T.~J.}\ \bibnamefont {Weiler}},\ }\href {\doibase
  10.1103/PhysRevD.101.015012} {\bibfield  {journal} {\bibinfo  {journal}
  {Phys. Rev. D}\ }\textbf {\bibinfo {volume} {101}},\ \bibinfo {pages}
  {015012} (\bibinfo {year} {2020})},\ \Eprint
  {http://arxiv.org/abs/1905.00046} {arXiv:1905.00046 [hep-ph]} \BibitemShut
  {NoStop}%
\bibitem [{\citenamefont {Essig}\ \emph {et~al.}(2020)\citenamefont {Essig},
  \citenamefont {Pradler}, \citenamefont {Sholapurkar},\ and\ \citenamefont
  {Yu}}]{Essig:2019xkx}%
  \BibitemOpen
  \bibfield  {author} {\bibinfo {author} {\bibfnamefont {R.}~\bibnamefont
  {Essig}}, \bibinfo {author} {\bibfnamefont {J.}~\bibnamefont {Pradler}},
  \bibinfo {author} {\bibfnamefont {M.}~\bibnamefont {Sholapurkar}}, \ and\
  \bibinfo {author} {\bibfnamefont {T.-T.}\ \bibnamefont {Yu}},\ }\href
  {\doibase 10.1103/PhysRevLett.124.021801} {\bibfield  {journal} {\bibinfo
  {journal} {Phys. Rev. Lett.}\ }\textbf {\bibinfo {volume} {124}},\ \bibinfo
  {pages} {021801} (\bibinfo {year} {2020})},\ \Eprint
  {http://arxiv.org/abs/1908.10881} {arXiv:1908.10881 [hep-ph]} \BibitemShut
  {NoStop}%
\bibitem [{\citenamefont {Baxter}\ \emph {et~al.}(2020)\citenamefont {Baxter},
  \citenamefont {Kahn},\ and\ \citenamefont {Krnjaic}}]{Baxter:2019pnz}%
  \BibitemOpen
  \bibfield  {author} {\bibinfo {author} {\bibfnamefont {D.}~\bibnamefont
  {Baxter}}, \bibinfo {author} {\bibfnamefont {Y.}~\bibnamefont {Kahn}}, \ and\
  \bibinfo {author} {\bibfnamefont {G.}~\bibnamefont {Krnjaic}},\ }\href
  {\doibase 10.1103/PhysRevD.101.076014} {\bibfield  {journal} {\bibinfo
  {journal} {Phys. Rev. D}\ }\textbf {\bibinfo {volume} {101}},\ \bibinfo
  {pages} {076014} (\bibinfo {year} {2020})},\ \Eprint
  {http://arxiv.org/abs/1908.00012} {arXiv:1908.00012 [hep-ph]} \BibitemShut
  {NoStop}%
\bibitem [{\citenamefont {Chen}\ \emph {et~al.}(2017)\citenamefont {Chen},
  \citenamefont {Chi}, \citenamefont {Liu},\ and\ \citenamefont
  {Wu}}]{Chen:2016eab}%
  \BibitemOpen
  \bibfield  {author} {\bibinfo {author} {\bibfnamefont {J.-W.}\ \bibnamefont
  {Chen}}, \bibinfo {author} {\bibfnamefont {H.-C.}\ \bibnamefont {Chi}},
  \bibinfo {author} {\bibfnamefont {C.-P.}\ \bibnamefont {Liu}}, \ and\
  \bibinfo {author} {\bibfnamefont {C.-P.}\ \bibnamefont {Wu}},\ }\href
  {\doibase 10.1016/j.physletb.2017.10.029} {\bibfield  {journal} {\bibinfo
  {journal} {Phys. Lett. B}\ }\textbf {\bibinfo {volume} {774}},\ \bibinfo
  {pages} {656} (\bibinfo {year} {2017})},\ \Eprint
  {http://arxiv.org/abs/1610.04177} {arXiv:1610.04177 [hep-ex]} \BibitemShut
  {NoStop}%
\bibitem [{\citenamefont {{Landau}}\ and\ \citenamefont
  {{Lifshitz}}(1991)}]{Landau_Lifshitz:1991qm}%
  \BibitemOpen
  \bibfield  {author} {\bibinfo {author} {\bibfnamefont {L.~D.}\ \bibnamefont
  {{Landau}}}\ and\ \bibinfo {author} {\bibfnamefont {E.~M.}\ \bibnamefont
  {{Lifshitz}}},\ }\href@noop {} {\emph {\bibinfo {title} {{Quantum Mechanics
  (Non-relativistic Theory)}}}}\ (\bibinfo  {publisher} {Butterworth
  Heinemann},\ \bibinfo {address} {Oxford},\ \bibinfo {year}
  {1991})\BibitemShut {NoStop}%
\bibitem [{\citenamefont {Chen}\ \emph
  {et~al.}(2015{\natexlab{a}})\citenamefont {Chen}, \citenamefont {Chi},
  \citenamefont {Liu}, \citenamefont {Wu},\ and\ \citenamefont
  {Wu}}]{Chen:2015pha}%
  \BibitemOpen
  \bibfield  {author} {\bibinfo {author} {\bibfnamefont {J.-W.}\ \bibnamefont
  {Chen}}, \bibinfo {author} {\bibfnamefont {H.-C.}\ \bibnamefont {Chi}},
  \bibinfo {author} {\bibfnamefont {C.~P.}\ \bibnamefont {Liu}}, \bibinfo
  {author} {\bibfnamefont {C.-L.}\ \bibnamefont {Wu}}, \ and\ \bibinfo {author}
  {\bibfnamefont {C.-P.}\ \bibnamefont {Wu}},\ }\href {\doibase
  10.1103/PhysRevD.92.096013} {\bibfield  {journal} {\bibinfo  {journal} {Phys.
  Rev. D}\ }\textbf {\bibinfo {volume} {92}},\ \bibinfo {pages} {096013}
  (\bibinfo {year} {2015}{\natexlab{a}})},\ \Eprint
  {http://arxiv.org/abs/1508.03508} {arXiv:1508.03508 [hep-ph]} \BibitemShut
  {NoStop}%
\bibitem [{\citenamefont {Siegert}(1939)}]{Siegert:1939zz}%
  \BibitemOpen
  \bibfield  {author} {\bibinfo {author} {\bibfnamefont {A.}~\bibnamefont
  {Siegert}},\ }\href {\doibase 10.1103/PhysRev.56.750} {\bibfield  {journal}
  {\bibinfo  {journal} {Phys. Rev.}\ }\textbf {\bibinfo {volume} {56}},\
  \bibinfo {pages} {750} (\bibinfo {year} {1939})}\BibitemShut {NoStop}%
\bibitem [{\citenamefont {Lin}(1977)}]{Lin:1977dl}%
  \BibitemOpen
  \bibfield  {author} {\bibinfo {author} {\bibfnamefont {D.~L.}\ \bibnamefont
  {Lin}},\ }\href {\doibase 10.1103/PhysRevA.16.600} {\bibfield  {journal}
  {\bibinfo  {journal} {Phys. Rev. A}\ }\textbf {\bibinfo {volume} {16}},\
  \bibinfo {pages} {600} (\bibinfo {year} {1977})}\BibitemShut {NoStop}%
\bibitem [{\citenamefont {Grant}(1974)}]{Grant_1974}%
  \BibitemOpen
  \bibfield  {author} {\bibinfo {author} {\bibfnamefont {I.~P.}\ \bibnamefont
  {Grant}},\ }\href {\doibase 10.1088/0022-3700/7/12/007} {\bibfield  {journal}
  {\bibinfo  {journal} {J. Phys. B}\ }\textbf {\bibinfo {volume} {7}},\
  \bibinfo {pages} {1458} (\bibinfo {year} {1974})}\BibitemShut {NoStop}%
\bibitem [{\citenamefont {{West}}\ and\ \citenamefont
  {{Morton}}(1978)}]{West:1978xe}%
  \BibitemOpen
  \bibfield  {author} {\bibinfo {author} {\bibfnamefont {J.~B.}\ \bibnamefont
  {{West}}}\ and\ \bibinfo {author} {\bibfnamefont {J.}~\bibnamefont
  {{Morton}}},\ }\href {\doibase 10.1016/0092-640X(78)90010-4} {\bibfield
  {journal} {\bibinfo  {journal} {At. Data Nucl. Data Tables}\ }\textbf
  {\bibinfo {volume} {22}},\ \bibinfo {pages} {103} (\bibinfo {year}
  {1978})}\BibitemShut {NoStop}%
\bibitem [{\citenamefont {{Henke}}\ \emph {et~al.}(1993)\citenamefont
  {{Henke}}, \citenamefont {{Gullikson}},\ and\ \citenamefont
  {{Davis}}}]{Henke:1993gd}%
  \BibitemOpen
  \bibfield  {author} {\bibinfo {author} {\bibfnamefont {B.~L.}\ \bibnamefont
  {{Henke}}}, \bibinfo {author} {\bibfnamefont {E.~M.}\ \bibnamefont
  {{Gullikson}}}, \ and\ \bibinfo {author} {\bibfnamefont {J.~C.}\ \bibnamefont
  {{Davis}}},\ }\href {\doibase 10.1006/adnd.1993.1013} {\bibfield  {journal}
  {\bibinfo  {journal} {Atom. Data Nucl. Data Tabl.}\ }\textbf {\bibinfo
  {volume} {54}},\ \bibinfo {pages} {181} (\bibinfo {year} {1993})}\BibitemShut
  {NoStop}%
\bibitem [{\citenamefont {Samson}\ and\ \citenamefont
  {Stolte}(2002)}]{Samson:2002xe}%
  \BibitemOpen
  \bibfield  {author} {\bibinfo {author} {\bibfnamefont {J.}~\bibnamefont
  {Samson}}\ and\ \bibinfo {author} {\bibfnamefont {W.}~\bibnamefont
  {Stolte}},\ }\href {\doibase http://dx.doi.org/10.1016/S0368-2048(02)00026-9}
  {\bibfield  {journal} {\bibinfo  {journal} {J. Electron Spectrosc. Relat.
  Phenom.}\ }\textbf {\bibinfo {volume} {123}},\ \bibinfo {pages} {265 }
  (\bibinfo {year} {2002})}\BibitemShut {NoStop}%
\bibitem [{\citenamefont {Suzuki}\ and\ \citenamefont
  {Saito}(2003)}]{Suzuki:2003xe}%
  \BibitemOpen
  \bibfield  {author} {\bibinfo {author} {\bibfnamefont {I.~H.}\ \bibnamefont
  {Suzuki}}\ and\ \bibinfo {author} {\bibfnamefont {N.}~\bibnamefont {Saito}},\
  }\href {\doibase http://dx.doi.org/10.1016/S0368-2048(03)00012-4} {\bibfield
  {journal} {\bibinfo  {journal} {J. Electron Spectrosc. Relat. Phenom.}\
  }\textbf {\bibinfo {volume} {129}},\ \bibinfo {pages} {71 } (\bibinfo {year}
  {2003})}\BibitemShut {NoStop}%
\bibitem [{\citenamefont {Zheng}\ \emph {et~al.}(2006)\citenamefont {Zheng},
  \citenamefont {Cui}, \citenamefont {Zhao}, \citenamefont {Zhao},\ and\
  \citenamefont {Chen}}]{Zheng:2006xe}%
  \BibitemOpen
  \bibfield  {author} {\bibinfo {author} {\bibfnamefont {L.}~\bibnamefont
  {Zheng}}, \bibinfo {author} {\bibfnamefont {M.}~\bibnamefont {Cui}}, \bibinfo
  {author} {\bibfnamefont {Y.}~\bibnamefont {Zhao}}, \bibinfo {author}
  {\bibfnamefont {J.}~\bibnamefont {Zhao}}, \ and\ \bibinfo {author}
  {\bibfnamefont {K.}~\bibnamefont {Chen}},\ }\href {\doibase
  http://dx.doi.org/10.1016/j.elspec.2006.04.006} {\bibfield  {journal}
  {\bibinfo  {journal} {J. Electron Spectrosc. Relat. Phenom.}\ }\textbf
  {\bibinfo {volume} {152}},\ \bibinfo {pages} {143 } (\bibinfo {year}
  {2006})}\BibitemShut {NoStop}%
\bibitem [{\citenamefont {{Gu}}(2008)}]{Gu:2008fac}%
  \BibitemOpen
  \bibfield  {author} {\bibinfo {author} {\bibfnamefont {M.~F.}\ \bibnamefont
  {{Gu}}},\ }\href {\doibase 10.1139/P07-197} {\bibfield  {journal} {\bibinfo
  {journal} {Can. J. of Phys.}\ }\textbf {\bibinfo {volume} {86}},\ \bibinfo
  {pages} {675} (\bibinfo {year} {2008})}\BibitemShut {NoStop}%
\bibitem [{\citenamefont {Lewin}\ and\ \citenamefont
  {Smith}(1996)}]{Lewin:1995rx}%
  \BibitemOpen
  \bibfield  {author} {\bibinfo {author} {\bibfnamefont {J.}~\bibnamefont
  {Lewin}}\ and\ \bibinfo {author} {\bibfnamefont {P.}~\bibnamefont {Smith}},\
  }\href {\doibase 10.1016/S0927-6505(96)00047-3} {\bibfield  {journal}
  {\bibinfo  {journal} {Astropart.Phys.}\ }\textbf {\bibinfo {volume} {6}},\
  \bibinfo {pages} {87} (\bibinfo {year} {1996})}\BibitemShut {NoStop}%
\bibitem [{\citenamefont {Sze}(1981)}]{Sze:1981sem}%
  \BibitemOpen
  \bibfield  {author} {\bibinfo {author} {\bibfnamefont {S.~M.}\ \bibnamefont
  {Sze}},\ }\href@noop {} {\emph {\bibinfo {title} {Physics of Semiconductor
  Devices}}}\ (\bibinfo  {publisher} {John Wiley and Sons},\ \bibinfo {address}
  {N. Y.},\ \bibinfo {year} {1981})\BibitemShut {NoStop}%
\bibitem [{\citenamefont {Philipp}\ and\ \citenamefont
  {Taft}(1959)}]{Philipp:1959ge}%
  \BibitemOpen
  \bibfield  {author} {\bibinfo {author} {\bibfnamefont {H.~R.}\ \bibnamefont
  {Philipp}}\ and\ \bibinfo {author} {\bibfnamefont {E.~A.}\ \bibnamefont
  {Taft}},\ }\href {\doibase 10.1103/physrev.113.1002} {\bibfield  {journal}
  {\bibinfo  {journal} {Phys. Rev.}\ }\textbf {\bibinfo {volume} {113}},\
  \bibinfo {pages} {1002} (\bibinfo {year} {1959})}\BibitemShut {NoStop}%
\bibitem [{\citenamefont {Grilli~di Cortona}\ \emph {et~al.}(2020)\citenamefont
  {Grilli~di Cortona}, \citenamefont {Messina},\ and\ \citenamefont
  {Piacentini}}]{GrillidiCortona:2020owp}%
  \BibitemOpen
  \bibfield  {author} {\bibinfo {author} {\bibfnamefont {G.}~\bibnamefont
  {Grilli~di Cortona}}, \bibinfo {author} {\bibfnamefont {A.}~\bibnamefont
  {Messina}}, \ and\ \bibinfo {author} {\bibfnamefont {S.}~\bibnamefont
  {Piacentini}},\ }\href@noop {} {\  (\bibinfo {year} {2020})},\ \Eprint
  {http://arxiv.org/abs/2006.02453} {arXiv:2006.02453 [hep-ph]} \BibitemShut
  {NoStop}%
\bibitem [{\citenamefont {Chen}\ \emph {et~al.}(2014)\citenamefont {Chen},
  \citenamefont {Chi}, \citenamefont {Huang}, \citenamefont {Liu},
  \citenamefont {Shiao} \emph {et~al.}}]{Chen:2013lba}%
  \BibitemOpen
  \bibfield  {author} {\bibinfo {author} {\bibfnamefont {J.-W.}\ \bibnamefont
  {Chen}}, \bibinfo {author} {\bibfnamefont {H.-C.}\ \bibnamefont {Chi}},
  \bibinfo {author} {\bibfnamefont {K.-N.}\ \bibnamefont {Huang}}, \bibinfo
  {author} {\bibfnamefont {C.-P.}\ \bibnamefont {Liu}}, \bibinfo {author}
  {\bibfnamefont {H.-T.}\ \bibnamefont {Shiao}},  \emph {et~al.},\ }\href
  {\doibase 10.1016/j.physletb.2014.02.036} {\bibfield  {journal} {\bibinfo
  {journal} {Phys. Lett. B}\ }\textbf {\bibinfo {volume} {731}},\ \bibinfo
  {pages} {159} (\bibinfo {year} {2014})},\ \Eprint
  {http://arxiv.org/abs/1311.5294} {arXiv:1311.5294 [hep-ph]} \BibitemShut
  {NoStop}%
\bibitem [{\citenamefont {Chen}\ \emph
  {et~al.}(2015{\natexlab{b}})\citenamefont {Chen}, \citenamefont {Chi},
  \citenamefont {Huang}, \citenamefont {Li}, \citenamefont {Liu}, \citenamefont
  {Singh}, \citenamefont {Wong}, \citenamefont {Wu},\ and\ \citenamefont
  {Wu}}]{Chen:2014ypv}%
  \BibitemOpen
  \bibfield  {author} {\bibinfo {author} {\bibfnamefont {J.-W.}\ \bibnamefont
  {Chen}}, \bibinfo {author} {\bibfnamefont {H.-C.}\ \bibnamefont {Chi}},
  \bibinfo {author} {\bibfnamefont {K.-N.}\ \bibnamefont {Huang}}, \bibinfo
  {author} {\bibfnamefont {H.-B.}\ \bibnamefont {Li}}, \bibinfo {author}
  {\bibfnamefont {C.-P.}\ \bibnamefont {Liu}}, \bibinfo {author} {\bibfnamefont
  {L.}~\bibnamefont {Singh}}, \bibinfo {author} {\bibfnamefont {H.~T.}\
  \bibnamefont {Wong}}, \bibinfo {author} {\bibfnamefont {C.-L.}\ \bibnamefont
  {Wu}}, \ and\ \bibinfo {author} {\bibfnamefont {C.-P.}\ \bibnamefont {Wu}},\
  }\href {\doibase 10.1103/PhysRevD.91.013005} {\bibfield  {journal} {\bibinfo
  {journal} {Phys. Rev. D}\ }\textbf {\bibinfo {volume} {91}},\ \bibinfo
  {pages} {013005} (\bibinfo {year} {2015}{\natexlab{b}})},\ \Eprint
  {http://arxiv.org/abs/1411.0574} {arXiv:1411.0574 [hep-ph]} \BibitemShut
  {NoStop}%
\end{thebibliography}%

\end{document}